\documentclass[11pt,twoside]{article}
\usepackage{asp2010}

\resetcounters

\begin{document}

\title{Multi-wavelength observations of Isolated Neutron Stars}
\author{R.P. Mignani$^{1,2}$ 
\affil{$^1$Mullard Space Science Laboratory, University College London, Holmbury St. Mary, Dorking, Surrey, RH5 6NT, UK}
\affil{$^2$Kepler Institute of Astronomy, University of Zielona G\'ora, Lubuska 2, 65-265, Zielona G\'ora, Poland}
}

\begin{abstract}
Almost 30 Isolated Neutron Stars (INSs) of different flavours have been identified at optical, ultraviolet, or infrared (UVOIR) wavelengths. Here, I present a short review of the historical background and describe the scientific impact of INS observations in the UVOIR.  Then, I focus on UVOIR observations of rotation-powered pulsars, so far the most numerous class of INSs identified at these wavelengths, and their observational properties. Finally, I present the results of new UVOIR observations and an update of the  follow-ups of $\gamma$-ray pulsars detected by {\em Fermi}. 
\end{abstract}

\section{Historical background}

Optical observations of pulsars started soon after their discovery, with B1919+21 being the first observed in search for its optical counterpart  (Ryle \& Bailey 1968). Soon after came the identification of the Crab pulsar (V=16.5) through the detection of optical pulsations
(Cocke et al.\ 1969a).  Searches for the optical counterpart  of the next best target, the Vela pulsar, were performed, albeit with no success (e.g., Cocke et al.\ 1969b).
It was only a few years later that its optical counterpart (V=23.6) was discovred (Lasker 1976) and the identification confirmed by the detection of optical pulsations (Wallace et al.\ 1977).  The third, and for a long time the last, 
optical pulsar to be discovered was  B0540$-$69 (V=22.5)  in the LMC (Middleditch \& Pennypacker 1985),  although its counterpart was unresolved from the  SNR\, 0540$-$69. The turn-off point  came at the end of the 1980s with the advent of CCD technology and the deployment of new generation 4m-class telescopes, like the ESO NTT in 1989. The NTT identified the B0540$-$69 counterpart and  those of the  two middle-aged (0.1--0.3 Myr) pulsars Geminga, and   B0656+14 (Caraveo et al.\ 1992,1994a,b; Bignami et al.\ 1993). The launch of the {\em HST} in 1990 and the deployment of the four 8.4m units of the ESO VLT array in 2000s represented a major burst for pulsar optical astronomy.  In particular, the {\em HST} made it possible observe pulsars in the near-ultraviolet (nUV) for the first time and  in the near-infrared (nIR) without being limited by the atmosphere, and, thanks to its sharper angular resolution,  to measure proper motion and parallax for both radio loud and radio silent pulsars, like Geminga. On the other hand, thanks to its large collecting power, the VLT  allowed to perform spectroscopy and polarimetry observations for pulsars much fainter than the Crab.  {\em HST} and VLT observations, as well as with other 8m-class telescopes, have been also instrumental to study the nature of newly-discovered classes of isolated neutron stars (INSs), such as the Soft Gamma-ray Repeaters (SGRs), the Anomalous X-ray Pulsars (AXPs), the  Central Compact Objects (CCOs), and the Rotating RAdio Transients  (RRATs).  So far, 26 INSs have been identified in the UV, optical, IR (UVOIR), plus 3 possible candidates (Mignani 2011). Together, the {\em HST} and VLT identified the majority of INSs detected in the UVOIR, studied most of them, and searched for the counterparts of many others (e.g., Mignani 2009, 2010a).

 \section{Scientific impact}

UVOIR observations are important in many fields of neutron star astronomy. Multi-band photometry and spectroscopy are key for characterising the neutron star spectral energy distribution over 10 energy decades and understanding the physics of the radiation processes in pulsar magnetospheres. In particular, UVOIR observations unveil spectral breaks in the magnetospheric emission at low energies. The characterisation of the broken-power law which fits the the optical--to--$\gamma$-ray spectrum, then allows one to track the particle energy and density distribution in the magnetosphere. Studying the neutron star UVOIR light curves, together with the X-ray and $\gamma$-ray ones, completes the mapping of the emission regions in the magnetosphere,  while  the measurement of polarisation, so far mostly feasible in radio and in the UVOIR, is crucial both to test magnetosphere models and determine whether, and how much, the atmosphere is magnetised. UVOIR timing also allows one to search for Giant Pulses, so far detected simultaneously only in radio and in the optical, and study the link between coherent and incoherent emission (Shearer et al., these proc.). Optical-UV observations are crucial for building the thermal map on the neutron star surface and unveil anisotropies in the temperature distribution, with X-ray and optical-UV emission produced by regions at different temperature.  Moreover, they allow one to test the decay of neutron star cooling curves for ages $>$1 Myr, where X-rays are detected only from small hot spots or polar caps and the bulk of the thermal emission is detected below 0.1 keV.   Optical-IR observations are important to search for debris disks, hinted by a black-body like hump in the IR spectrum, and by systematic phase lags and smearing of the optical-IR light curve with respect to X-ray one. On a larger scale, the search for bow-shocks undetected in radio/X-rays allows one to  constrain the pulsar radial velocity by fitting the bow-shock position angle with respect to the plane of the sky. The inferred space velocity, with the tangential velocity derived from the proper motion and distance, then makes it possible to back-ward extrapolate the pulsar orbit in the Galaxy and investigate parental cluster association (e.g., Mignani et al. 2012a). On the other hand, thanks to their better spatial resolution, optical observations allows one to resolve the structure of pulsar-wind nebulae (PWNe) unresolved with  {\em Chandra} such as, the   B0540$-$69 nebula (De Luca et al.\ 2007).   {\em HST} observations
are also  key in the investigation of the mysterious $\gamma$-ray flares from the Crab Nebula (Tavani et al.\ 2011). Last but not least, optical observations allowed to measure proper motion, parallaxes, and absolute positions for radio-silent  or radio-faint pulsars, e.g. the Crab (Kaplan et al.\ 2008), Vela (Caraveo et al.\ 2001),   Geminga (Caraveo et al.\ 1996), B0540$-$69,   B1055$-$52 (Mignani et al.\ 2010a,b).

\section{UVOIR emission properties of rotation-powered pulsars}

Ten rotation-powered pulsars (RPPs), plus 3 candidates, are identified in the UVOIR.  They are all detected in the optical, with only some also in the nUV and nIR. The knowledge of the spectrum mostly relies on broad-band photometry, with  spectroscopy is available for 5 RPPs only.  In young ($<$ 10 kyr) RPPs, the spectrum is usually dominated by a flattish power-law (PL) continuum $F_{\nu} \propto \nu^{-\alpha_{O}}$ (e.g. Mignani et al.\ 2007), ascribed to synchrotron radiation. The optical luminosity $L_{O}$  scales with the rotational energy $E_{rot}$, as  $L_{O} \propto E_{rot}^{(1.70\pm0.03)}$. The  comparison with similar relations for the X-ray  $L_{X} \propto E_{rot}^{(1.04\pm0.09)}$ (Marelli et al.\ 2011) and $\gamma$-ray $L_{\gamma} \propto E_{rot}^
{0.5}$ (Abdo et al.\ 2010) luminosities implies a steeper dependence on $E_{rot}$ for the optical than the other bands. The optical luminosity is also predicted to scale as a function of the magnetic field $B$ and spin period $P$ as $L_{\rm opt} \propto B^{4}P^{-2}$ (Pacini\&Salvati 1983), which implies a secular decrease, only tentatively measured for the Crab pulsar  ($8\pm4$ mmag/yr; Nasuti et al.\ 1996; 2.9$\pm$1.6 mmag/yr; Sandberg \& Sollerman 2009).  The optical emission efficiency $\eta_{\rm opt}= L_{\rm opt}/E_{\rm rot}$ decreases sharply for ages of $\sim$10 kyrs  and remains more or less constant for older RPPs ($>$ 0.1 Myrs). The optical spectral index $\alpha_{O}$ possibly evolves with  the spin-down age, but there is no correlation with the X-ray spectral index $\alpha_{X}$, and, generally, $\alpha_{O}<\alpha_{X}$, which implies spectral breaks (Mignani et al.\ 2010a) and flattening of the PL spectrum between the X-ray and optical bands. For middle-aged RPPs (0.1--1 Myr), the optical spectrum features an additional Rayleigh-Jeans (R-J) component, with temperatures $T\sim 0.3$--0.7 MK (Mignani et al.\ 2010b; Kargaltsev \& Pavlov 2007). Possible evidence of spectral absorption or emission features on the PL+R-J continuum, originally suggested also for Geminga (Bignami et al.\ 1996; Mignani et al.\ 1998), has been  found for   B0656+14 (Durant et al.\ 2011). For older RPPs, the spectrum is still poorly determined.
NUV emission  has been detected for 8 RPPs. For the  Crab and Vela pulsars, the nUV spectrum follows a PL, with a small change of the PL slope in the optical/nUV for the former. For middle-aged RPPs, the nUV spectrum is dominated by the  R-J continuum observed in the optical.  For   B1055$-$52, the nUV R-J is above the extrapolation of the two blackbody (BB) components to the X-ray spectrum (Mignani et al.\ 2010b),  suggesting a three-component thermal spectrum, unlike Geminga and   B0656+14, where the nUV R-J is more compatible with the extrapolation of the coldest X-ray BB.   For the 1.5 Gyr old  milli-second pulsar J0437$-$4715, the R-J temperature is $\sim$ 0.1 MK (Durant et al.\ 2012), well above cooling model predictions, suggesting a reheating processes of the neutron star interior.  Observations of RPPs in the nUV still rely on the {\em HST}, with {\em GALEX} not sensitive enough (Mignani et al. in prep). Observations in the  extreme UV (eUV) are crucial to fill the gap between the  optical and soft X-ray bands. Ten RPPs were observed by the {\em EUVE} (Korpela \& Bowyer 1998), but only Geminga,   B0656+14, and J0437$-$4715 were detected and, in all cases, the eUV spectrum was found consistent with a R-J.
NIR emission has been detected for 6 RPPs. After the Crab, the {\em HST}, VLT, and Magellan detected Geminga,   B0656+14 (Koptsevich et al.\ 2001), Vela (Shibanov et al.\ 2003), B1509$-$58 (Kaplan \& Moon 2006), and B0540$-$69 (Mignani et al.\ 2012b). In all cases, the nIR spectrum follows the optical PL extrapolation. Interestingly enough, the steep nIR spectrum of B0656+14 was initially associated with the presence of a debris disk (Perna et al.\ 2000). In the mid-IR, emission  both the Crab (Temim et al.\ 2009) and Vela pulsars, and possibly Geminga, have been detected (Danilenko et al.\ 2011) with {\em Spitzer}, while most of the other RPPs are undetected(Mignani et al, in prep.) Interestingly, for  Vela the mid-IR  PL spectral is much steeper than the nIR, which might either be due to a spectral break in the PL emission or the presence of a debris disk. At present, mid-IR observations of RPPs are limited by {\em Spitzer} being now in its ÒwarmÓ phase, while {\em Herschel} has not a sufficient  spatial resolution, and {\em JWST} is still a few years ahead.
Pulsations have been detected from 5 RPPs, all but B0540$-$69 both in the optical and nUV (Mignani 2010b, 2012), and only for the Crab pulsations in the nIR have been detected. Light Curves are double-peaked, not always aligned with the $\gamma$/X/radio peaks. nUV observations of the Crab showed the dependence of its light curve on wavelength, with both the  peak widths and separation larger in the optical than in the nUV (Percival et al.\ 1993), possibly associated with the change of the PL slope in the optical-nUV. Indeed, there is no clear dependence on wavelength in the nUV (Gull et al.\ 1998), where the spectral index is constant. Linear optical polarisation has been measured for 4 RPPs, but only for the Crab (S\l{}owikowska et al.\ 2009) phase-resolved measurements across the whole phase exist, while for B0656+14 they only cover  30\% of the phase (Kern et al.\ 2003). For both Vela and B0540$-$69 (Mignani et al.\ 2007b; 2010a) only phase-integrated measurements exist. For the Crab, the polarisation degree is maximum in the inter-pulse and the polarisation angle swings before the peaks, with the DC component possibly due to the emission knot $0\farcs6$ away from the pulsar (S\l{}owikowska et al.; Moran et al., these proc.). In all cases, the phase-averaged polarisation degree is $\la 10\%$ and, at least for the Crab and Vela pulsars, the polarisation direction is nearly aligned with the pulsar proper motion vector and the symmetry axis of the PWN.

\section{Recent UVOIR observations}

B0540$-$69 was detected in the nIR for the first time by the VLT with J=20.14, H=19.33, K=18.55 (Mignani et al.\ 2012a), and is the sixth RPPs detected in the nIR. Its optical/nIR spectrum is fit by a single PL with spectral index  $\alpha{O,nIR} = 0.70\pm0.04$, which implies no PL  break in the optical/nIR, although a double break still expected in the UV  (Mignani et al.\ 2010a).  Like in the optical, the nIR luminosity $L_{nIR}$ scales with $E_{rot}$, as $L_{nIR} \propto E_{rot}^{(1.70\pm0.03)}$. B0540$-$69 is the most efficient RPPs in the nIR, like in the optical, with $\eta_{nIR} \sim 10^{-5}$, a factor 10 higher than its ÒtwinÓ, the Crab pulsar. The VLT also detected the 0540$-$69 PWN in the nIR for the first time, and is the fifth PWN detected in the nIR/mIR.  The PWN morphology is consistent with the optical, with no evidence for the bright knot detected by {\em HST} (De Luca et al.\ 2007), and the PWN shows no spectral variations on small scales,  like in the optical. On a larger scale the PWN spectrum is described by a PL with $\alpha_{O,nIR} = 0.56\pm0.03$, flatter than in the mIR where, however, the pulsar  is not resolved by {\em Spitzer} and its contribution to the PWN flux is unknown.  
The first optical observations of the double pulsar J0737$-$3039 were recently obtained by the {\em HST} in the ACS/HRC 425W and 606W filters. No counterpart was detected down to $m_{435W} = 27.0$ and $m_{606W} = 28.3$ (Ferraro et al.\ 2012). The optical ULs are consistent with the best-fit  {\em XMM-Newton} spectrum:  a PL (photon index $\Gamma \sim 3.3$) plus a hot BB ($kT \sim 134$ eV,  emitting radius $r \sim$ 100 m)  for PSR-A, and a cold BB ($kT \sim 32$ eV, $r \sim$15 km) for PSR-B. Since the rotational energy for PSR-A and PSR-B is $E_{rot}^{A} = 5 .9 \times 10^{33}$ erg s$^{-1}$ and $E_{rot}^B = 1 .7 \times 10^{30}$ erg s$^{?1}$, respectively, the upper limits do not rule out that optical emission from the system could  be entirely produced by PSR-A's magnetosphere, assuming a spectral break at $\sim$ 0.1 keV to be compatible with the expected optical emission from the cold BB of PSR-B. This would $\eta_{O} < 8 \times 10^{-8}$,  quite high for the $\sim$200 Myr-old PSR-A. Alternatively, optical emission from the system might be entirely produced by the cold BB of PSR-B or to a very cold BB from PSR-A with $kT<20$ eV.
The harvest of $\gamma$-ray pulsars detected by {\em Fermi} (Nolan et al.\ 2012) is a primary reference for optical observations of RPPs.  After a pilot survey of $\sim 12$ {\em Fermi} pulsars with mid-size telescopes at the La Palma Observatory (Collins et al.\ 2011), targeted observations were performed with 8m-class telescopes. The Vela-like pulsar J1357$-$6429 was observed with the VLT and a candidate counterpart detected with  $I\sim24.6$; $V>27$ $R>26.8$ (Mignani et al.\ 2011). The comparison between the {\em Chandra} and radio positions also yielded its first proper motion measurement $\mu=0.17\pm0.055$ \arcsec/yr. The identification was confirmed by Danilenko et al.\ (2012) who claimed a detection in the $V$ and $R$ bands after subtraction of nearby stars. However, their photometry ($V = 27.3$; $R = 25.52$) would imply an extremely steep ($\alpha_{O} =5$--6.5 spectrum for J1357$-$6429,   which casts doubts on the  identification. A second Vela-like pulsar, B1046$-$58, was observed by the VLT but no counterpart was detected at the {\em Chandra} position down to $V>27.6$ (Mignani et al.\ 2011) and $R>26.4$ (Razzano et al.\ 2012). The third Vela-like pulsar observed with the VLT is J1028$-$5819 (Mignani et al.\ 2012c) but, due to the presence of a bright star  (Star A; $V=13.1$) at $\sim 4\arcsec$ from the pulsar, no detection was obtained ($B>25.4$; $V>25.3$). Optical observations were also used to verify the tentative {\em Swift}/XRT X-ray identification (Marelli et al.\ 2011). Indeed, the Star A's colours from the GSC-2, 2MASS, and unpublished {\em Swift}/UVOT observations, suggest an F-type stars, with a possible X-ray-to-optical flux ratio $F_{X}/F_{V} \sim 10^{-3}$--$10^{-2}$, which implies that, actually, it could be the counterpart to the {\em Swift}/XRT source associated with J1028$-$5819.  However, new observations of the X-ray source with {\em Suzaku} suggest a PL spectrum  ($\Gamma_{X}=1.7\pm0.2$), which supports the identification with the pulsar and suggest that  any contribution of Star A to the X-ray flux is probably negligible.  These observations are compatible a possible spectral break between the optical and X-rays but not between X and $\gamma$-rays. The X-ray luminosity  $L_{X} = (9.5\pm5. 9) \times10^{31}$ erg s$^{-1}$ implies an emission efficiency $\eta_{X} =(1.13\pm0.70) 10^{-4}$, consistent with the observed trends (Marelli et al.\ 2011). The inferred limits on $\eta_{O} < 10^{-6}$ are also consistent with the efficiency of other RPPs. {\em Suzaku} observations also showed evidence for an extended structure ($\sim 6\arcmin$) around the pulsar, with a count rate of  $(5.3\pm0.5) \times 10^{-4}$ c s$^{-1}$ arcmin$^{-2}$ in the 0.2--10 keV energy range, possibly associated with a PWN. The count-rate is consistent with the expected size and luminosity of a $\sim 90$ kyr old pulsar, for PWN X-ray luminosity of $10^{-4}$--$10^{-2} E_{rot}$ (Kargaltsev \& Pavlov 2010).  The VLT also detected the companion star to the millisecond pulsar J0610$-$2100 (Pallanca et al.\ 2012), a black widow system with a low-mass companion ($M \sim0.02 M_{\odot}$). The companion ($R=25.6$)  shows a periodic modulation which well correlates with the orbital period ($\sim$9 hrs). The optical observations are compatible with a companion with $T_{eff} \sim 3500$, luminosity $\sim 0.0017 L_{\odot}$ and radius $\sim 0.14 R_{\odot}$.  
Observations of {\em Fermi} pulsars have been also performed with the Spanish 10.4m GTC.  For J0007+7303, the first radio-silent {\em Fermi} pulsar,  no counterpart was detected down to $r\sim27.6$ (De Luca et al.\ 2012), much deeper than previous limits of $R\sim 25.1$ (Halpern et al.\ 2004) and $R\sim25.9$ (Collins et al.\ 2011). J0205+6449  in the 3C\, 58 SNR was observed both by the GTC and the Gemini, and a likely counterpart was detected ($i \sim26.6$; Shearer et al.\ 2012), barely resolved from its PWN, and marginally detected also by the TNG.

\acknowledgements RPM thanks the SOC for financial support and the LOC for the exquisite organisation of the ERPM conference.


\end{document}